# Influential Factors in Increasing an Amazon products Sales Rank

Ben Chen[1], Rohit Mokashi[1], Mamata Khadka[1], Robert Reyes[1], Huthaifa I. Ashqar[1,2]

[1] University of Maryland, Baltimore County

[2] Arab American University

## Abstract

Amazon is the world's number one online retailer and has nearly every product a person could need along with a treasure trove of product reviews to help consumers make educated purchases. Companies want to find a way to increase their sales in a very crowded market, and using this data is key.  A very good indicator of how a product is selling is its sales rank; which is calculated based on all-time sales of a product where recent sales are weighted more than older sales. Using the data from the Amazon products and reviews we determined that the most influential factors in determining the sales rank of a product were the number of products Amazon showed that other customers also bought, the number of products Amazon showed that customers also viewed, and the price of the product. These results were consistent for the Digital Music category, the Office Products category, and the subcategory Holsters under Cell Phones and Accessories.

## Introduction

Amazon is one of the most famous American multinational technology companies that focus on e-commerce, cloud computing, digital streaming, and artificial intelligence. The company has been the top choice for many third-party sellers to sell products worldwide. Likewise, the consumers want to buy products from Amazon as they offer a wide variety of relatively less expensive products. Consumers all around the world can review and compare the products to help their decision to buy the products that they want. Amazon, including other e-commerce platforms publishes the details about the products and allows consumers to write reviews for the products. The product attributes such as price, brand, ranking of product, features as well as reviews from the previous buyers are the most important factors in determining the sales rate of the product.

As a company, we wanted to choose a product to sell on Amazon. In order to make the decision, we analyzed the Amazon review data. The goal of the company is to identify the most important attributes of the product so that the appropriate product can be selected and sold on Amazon.

To perform the data analysis, we downloaded Amazon product review data from May 1996 to October 10`8 from http://deepyeti.ucsd.edu/jianmo/amazon/index.html, a webpage maintained by Jianmo Ni (1). The dataset contained over 241.1 million product reviews and included 29 different product categories (1).

For this project, we analyzed three different categories of products including cell phones and accessories, digital music, and office products. We analyzed product attributes from the review data such as reviewer id, the id of the product, name of the reviewer, vote, rating, and review related information. We will use these to determine what are the most influential factors in determining the sales rank of a product.

**Literature Review**

We will focus on papers related to the sales rank of Amazon products. Much research on the Amazon Review data set has been focused on sentiment analysis using natural language processing, which is beyond the scope of this paper.

Li et al. (2016) results showed that a large number of reviews and high product ratings lead to a better sales rank. Their main investigation was to find how product category, answered questions, discount, and review usefulness affected the relationship between the customer reviews and sales rank, which all were shown to be significant. They categorized products as those that are easy to obtain quality information on (search products), and those that are difficult to obtain useful information on their quality until used for themselves (experience products). There were differences in the importance of each factor between the two categories. While the initial results seem obvious, we can further explore the direct relationship between the other variable and the rank directly. The category differences may come up depending on which Amazon category we use, or if we further break down the category by subcategory.

Although currently in preprint, Etumnu (2022) has shown results that show that the sales rank of a product is increased by the number of questions and answers the products page has. This will be another variable that we can test and compare results with Etumnu (2022). The analysis was done on ground coffee products, and it will be interesting to see if these results show the same for other products.

Zhang et al. (2011) created a ranking model and compared it to the actual sales rank from Amazon. While they used sentiment analysis, more suited to our purpose they also used the reviews helpfulness votes, and the review posting date in their model. Their model showed a good correlation from their rank to Amazon's sales rank but was done on only two small subcategories. We can further explore the helpfulness votes, and posting dates in relationship to sales rank, and see the results on other products and categories.

Prasad et al. (2017) made use of networks of products bought together to estimate a product's rank. They created the network with the also-bought data available in the Amazon product dataset. They also used the volume of reviews, percentage of fake reviews, total time a product has been able to have been reviewed, number of words, and a few natural language processing calculated variables. The advantage of this network method was that it worked when there were no reviews, and it could classify very high from very low ranked products. It also had a very high accuracy in predicting the sales rank of CDs and cell phones. We can use the also bought information for each product in our model of predicting the sales rank of a product and see if the relationship holds for other products and categories.

There has been much research on the Amazon review dataset, but mostly using natural language processing in the models. There is room to further explore how other variables can affect the sales rank of a product while incorporating previously found important variables into a new model and testing on other categories.

## Dataset

The dataset was downloaded from http://deepyeti.ucsd.edu/jianmo/amazon/index.html. It contains over 231.1 million product reviews. The reviews data spans from May 1996 – October 2018. This dataset includes reviews (ratings, text, helpfulness votes), product metadata (descriptions, category information, price, brand, and image features), and links (also viewed/also bought graphs). The website provides following files: entire raw review data (34 GB - all 233.1 million reviews), ratings only (6.7 GB - without reviews or metadata),  5-core (14.3

GB - contains subset of the data in which all users and items have at least 5 reviews), pre-category data (the review and product metadata for each category). We performed our analysis on Digital Music, Office Products, and Holsters (a subset of Cell Phones and Accessories) categories.

Following are the product review data columns:

| Key | Description |
|---|---|
| reviewerId | Id of the reviewer, e.g. A2SUAM1J3GNN3B |
| asin | Amazon Standard Identification Number of the product, e.g., 0000013714 |
| reviewerName | Name of the reviewer |
| vote | Helpful votes on the review |
| verified | If the review is verified |
| style | A dictionary of the product metadata, e.g., "Format" is "Hardcover" |
| reviewText | Text of the review |
| overall | Rating of the product |
| summary | Summary of the review |
| unixReviewTime | Time of the review (unix time) |
| reviewTime | Time of the review (raw) |

| image | Images that users post after they have received the product |
|---|---|

Following are the product metadata columns:

| Key | Description |
|---|---|
| asin | Amazon Standard Identification Number of the product, e.g., 0000013714 |
| title | Name of the product |
| feature | Bullet-point format features of the product |
| description | Description of the product |
| price | Price in US dollars (at time of crawl) |
| imageURL | URL of the product image |
| imageURLHighRes | URL of the high-resolution product image |
| related | Related products (also bought, also viewed, bought together, buy after viewing) |
| salesRank | Sales rank information |
| brand | Brand name of the product |
| categories | List of categories product belongs to |
| tech1 | The first technical detail table of the product |

| tech2 | The second technical detail table of the product |
|-------|--------------------------------------------------|
| similar | Similar product table |

## Methods

We explored a large dataset of Amazon reviews to see what we may learn. There are many data variables that can affect and interact with each other. Understanding these relationships was key in our analysis. Especially, our aim was to analyze what variables or attributes of the data, apart from the textual data, were affecting the rankings of the product in a particular category. We performed our analysis on Digital Music, Office Products, and Holsters (a subset of Cell Phones and Accessories) categories.

We had two different data (product metadata and review data) for each category we analyzed. Hence, we employed different data cleaning strategy. For product metadata, we performed following things:

- Remove unnecessary columns

- Only keep products that have rank in the correct string format

- Drop duplicate products

- Only keep products that have price in the correct string format

- Create new column also_buy_len using the number of also buy products

- Create new column also_view_len using the number of also view products

- Extract ranks as integers from strings in the rank column

- Extract prices as floats from strings in the price column

- Remove products with invalid price

One straightforward way to identify the outliers is to generate the box plot of the values.

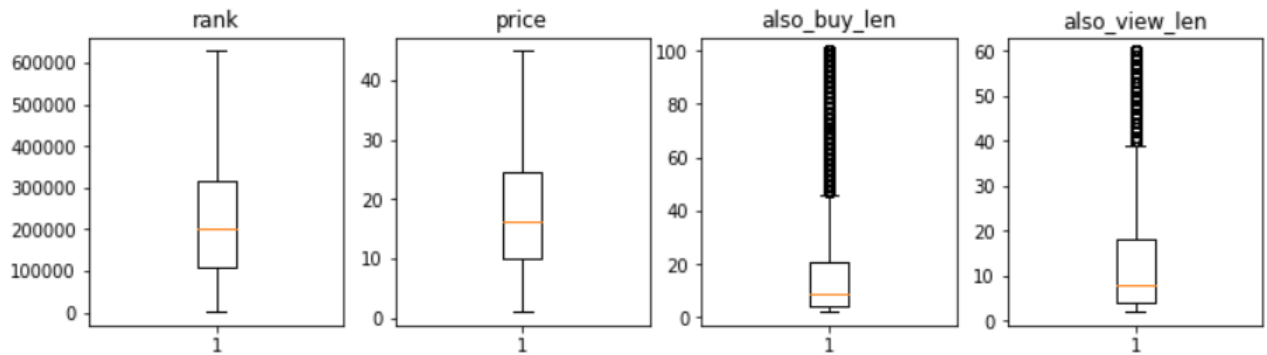

For product review data, we performed following things:

- Cast vote column to integer

- Remove reviews with missing vote

- Drop duplicate reviews

- Group reviews by product and calculate average overall rating, percentage of verified reviews, average number of helpful votes and number of reviews for each product

After the data wrangling step, we merged product metadata and product review data. We combined review columns and metadata columns with the help ASIN number of the products. After merging we created training and testing data with 30% test data.

This is the data obtained after combining review data and metadata. It contains following columns: price (price of the product), also_buy_len (count of also_buy), also_view_len (count of also_view), overall (mean of the overall ratings), verified (percentage of verified reviews), vote (mean and count of votes), count (count of reviews). The Reduced datadata contains price, also_buy_len, and also_view_len only. The normalized data is obtained after normalizing above data categories. We subtracted mean and divided by standard deviation to obtain the normalized values. Later, we scaled the values by multiplying by 100.

| | price | also_buy_len | also_view_len | overall | verified | vote | count |
|---|---|---|---|---|---|---|---|
| **price** | 1.000000 | 0.052659 | 0.085628 | 0.049514 | 0.034325 | 0.001707 | -0.027182 |
| **also_buy_len** | 0.052659 | 1.000000 | 0.643027 | 0.010134 | 0.175274 | 0.059815 | 0.195256 |
| **also_view_len** | 0.085628 | 0.643027 | 1.000000 | -0.083388 | 0.154485 | 0.066733 | 0.180891 |
| **overall** | 0.049514 | 0.010134 | -0.083388 | 1.000000 | -0.016967 | -0.117561 | -0.029645 |
| **verified** | 0.034325 | 0.175274 | 0.154485 | -0.016967 | 1.000000 | -0.167158 | 0.015951 |
| **vote** | 0.001707 | 0.059815 | 0.066733 | -0.117561 | -0.167158 | 1.000000 | 0.031781 |
| **count** | -0.027182 | 0.195256 | 0.180891 | -0.029645 | 0.015951 | 0.031781 | 1.000000 |

Above figure represents correlation between variables

In our analysis, we have used various regression methods for quantifying factor's impact. Ordinary Least Squares regression (OLS) is a popular method for calculating the coefficients of linear regression equations that describe the connection between one or more independent quantitative variables and a dependent variable (simple or multiple linear regression). Least squares stand for the minimum squares error (SSE). The OLS approach aims to reduce the sum of square differences between observed and predicted values.

Robust regression is a form of regression analysis designed to overcome some limitations of traditional parametric and non-parametric methods. Regression analysis seeks to find the relationship between one or more independent variables and a dependent variable. Robust regression methods are designed to be not overly affected by violations of assumptions by the underlying data-generating process. [Wiki]

Random Forest Regression is a supervised learning approach for regression that use the ensemble learning method. The ensemble learning method combines predictions from several machine learning algorithms to produce a more accurate prediction than a single model. The bootstrapping Random Forest approach combines ensemble learning methods with the decision tree framework to construct many randomly drawn decision trees from data, then averaging the results to produce a new result that frequently leads to high predictions/classifications.

We executed Random Forest regression model on Original Data, Reduced Data, Normalized Original Data, Normalized Reduced Data. For this we have used Scikit-learn, a standard Python library. The number of estimators (number of trees) and maximum depth (maximum dept of tree) we used for Original Data and reduced data were 500 and 5, respectively. For Normalized Data, we assigned maximum depth to 4.

| | Factor | Original Data | Normalized Data | Reduced Data |
|---|---|---|---|---|
| 1 | also_buy_len | 0.363361 | 0.440074 | 0.446031 |
| 2 | also_view_len | 0.175150 | 0.167970 | 0.250331 |
| 0 | price | 0.154402 | 0.125549 | 0.303638 |
| 6 | count | 0.118440 | 0.115771 | NaN |
| 4 | verified | 0.077387 | 0.077844 | NaN |
| 5 | vote | 0.067343 | 0.040028 | NaN |
| 3 | overall | 0.043919 | 0.032765 | NaN |

Above figure shows table containing factors arranged with ascending importance

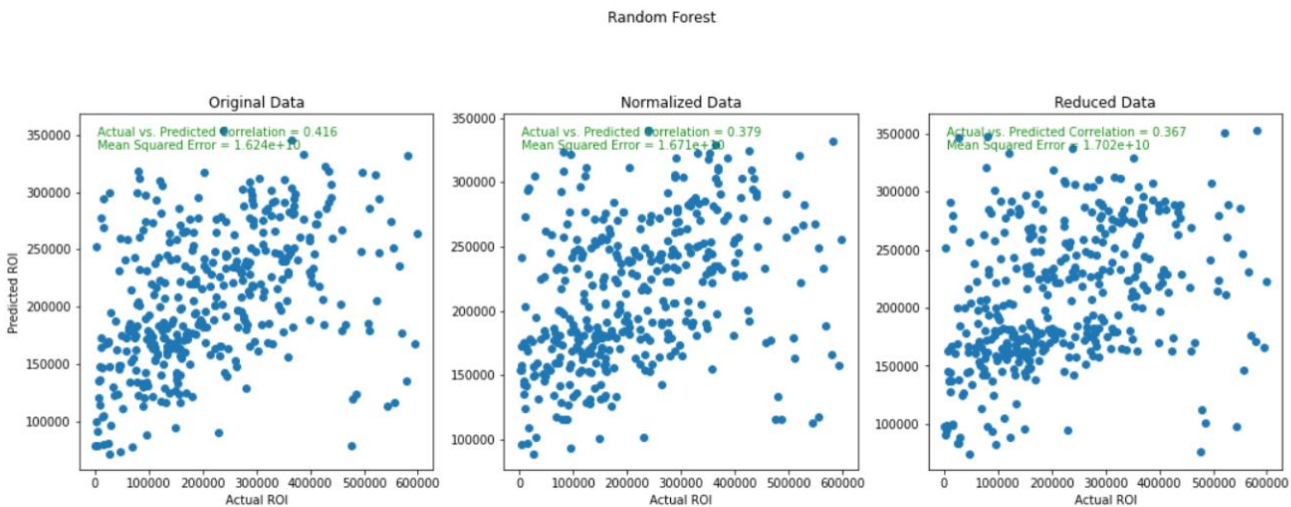

Above figure represents Actual Rank vs Predicted Rank graph.

K Nearest Neighbors

Both classification and regression issues can be solved with the KNN technique. The KNN algorithm predicts the values of new data points based on 'feature similarity.' This means that a value is assigned to the new point based on how similar it is to the points in the training set. KNN models are simple to use and effectively manage non-linearities. Model fitting is also usually quick. As Random Forest, we also executed KNN regression model on Original Data, Reduced Data, Normalized Original Data, Normalized Reduced Data. For this we have used Scikit-learn, a standard Python library. We passed number of neighbors parameter as 60.

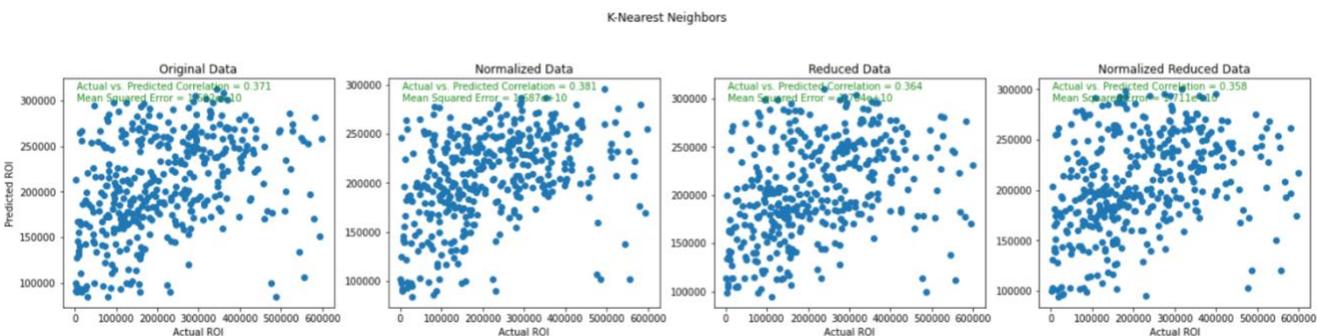

Naive Bayes methods are a set of supervised learning algorithms based on applying Bayes' theorem with the "naive" assumption of conditional independence between every pair of features given the value of the class variable. In Gaussian Naive Bayes algorithm, the likelihood of the features is assumed to be Gaussian. We executed Gaussian Naive Bayes model only on Original Data.

A decision tree is a supervised machine learning model used to predict a target by learning decision rules from features. Decision tree builds regression or classification models in the form of a tree structure. It breaks down a dataset into smaller and smaller subsets while at the same time an associated decision tree is incrementally developed. The result is a tree with decision nodes and leaf nodes. We executed Decision Tree regression model only on Original Data. For this we have used Scikit-learn, a standard Python library. We assigned 3 as maximum depth of the tree and assigned 0.3 to minimum impurity decrease parameter.

Multi-layer Perceptron (MLP) is a supervised learning algorithm that learns a function $f(.): R^m \rightarrow R^o$ by training on a dataset, where $m$ is the number of dimensions for input and $o$ is the number of dimensions for output. Given a set of features $X = x_1, x_2, \ldots, x_m$ and a target $y$, it can learn a non-linear function approximator for either classification or regression. We executed Multi-Layer Perceptron regression model only on Original Data. [https://scikit-learn.org/stable/modules/neural_networks_supervised.html] For this method, we assigned one to random state parameter which determines random number generation for weights and bias initialization. We kept maximum number of iterations to 6000.

The supervised learning algorithm Support Vector Regression is used to predict discrete values. SVMs and Support Vector Regression are both based on the same principle. SVR's primary concept is to identify the optimum fit line. The best fit line in SVR is the hyperplane with the greatest number of points. The SVR, unlike other regression models, aims to fit the best line within a threshold value, rather than minimizing the error between the real and predicted value. The distance between the hyperplane and the boundary line is the threshold value. We executed Support Vector regression model only on Original Data.

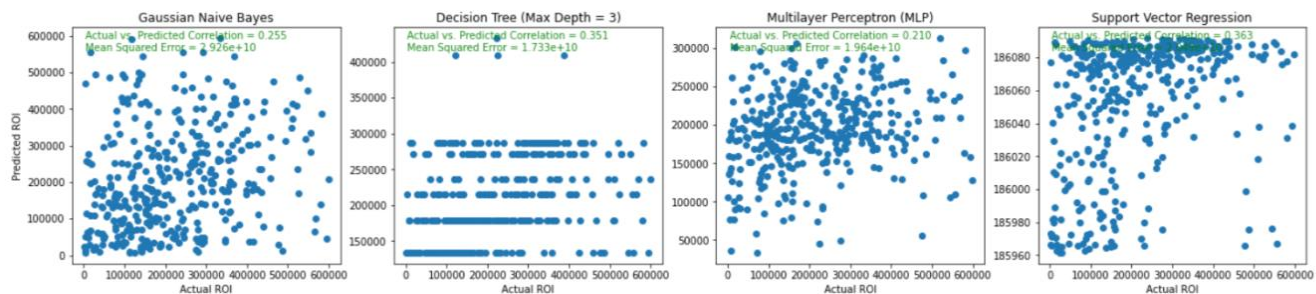

## Analysis and Results

In this section, we will focus on the results for the Digital Music category. Although the product rank of some other categories can be predicted with higher accuracy, all of them follow the same pattern.

The correlations between the independent variables are generally insignificant, except for the 0.64 correlation between *also_buy_len* and *also_view_len*. Despite their high correlation, removing either one of them will lower the out-of-sample prediction accuracy of our models. So, both variables are important, and we will keep both of them for our analysis.

We first want to get an idea of the predictive powers of our independent variables. We can indirectly get a sense of it by using the feature importance of the random forest model. The feature importance tells us how important each variable is for splitting nodes of a decision tree to reduce impurity. For our dataset, *also_buy_len* is the most important factor, followed by *also_view_len*, and *price*. All three of these variables are from the metadata. The rest of the variables, *count*, *verified*, *vote* and *overall* are much less important, and they are all from the reviews data. If we normalize the data by subtracting the mean and dividing by standard deviation, and retrain the model, the pattern of feature importance remains the same.

Next, we want to train different regression models and compare their out-of-sample performance. We mainly compare different models by comparing their mean squared errors (MSE), and correlations between actual product rank and predicted product rank.

For random forest, we trained our model in three different ways: using original data, using normalized data, and using only the top three variables in terms of feature importance (all from metadata). Random forest trained with original data gives the best result, with MSE of $1.62 \times 10^{10}$ and correlation of 0.416. Random forest trained with normalized data gives MSE of $1.67 \times 10^{10}$ and correlation of 0.379. This shows that we lost information when normalizing the data, and the random forest algorithm doesn't require data normalization. Random forest trained with only top three variables gives MSE of $1.70 \times 10^{10}$ and correlation of 0.367. This shows that those less important variables still provide predictive power, and our model becomes worse if we ignore those variables.

For linear regression, we trained our model in four different ways (all without normalization): ordinary least square (OLS), OLS with only top three variables, linear regression using Huber loss function (robust linear regression), linear regression using Huber loss function with only top three variables. The corresponding MSE are $1.68 \times 10^{10}$, $1.72 \times 10^{10}$, $1.70 \times 10^{10}$ and $1.75 \times 10^{10}$. The corresponding correlations are 0.383, 0.351, 0.386 and 0.355. Again, the result shows variables from the review data provide predictive value. The OLS coefficients are all significant if we use 90% confidence level. Only *vote* becomes insignificant if we use 95% confidence level. Coefficients of *price* and *count* are positive, meaning that predicted product ranks are worse when prices and number of reviews are higher. For *price*, it is easily understandable that lower price products have better ranks. For the number of reviews, it is possible that some bad products tend to have a lot of negative reviews, whereas average products don't get that many reviews. Coefficients of *also_buy_len*, *also_view_len*, *overall*, *verified* and *vote* are all negative, meaning that the higher these variables are, the better the predicted rank is. The above result should be consistent with our intuition. Products with higher *also_buy_len* and *also_view_len* are relatively popular products. Products with high *overall* are products with good ratings. Products with high *verified* are products with a lot of verified reviews (so they are usually not fake reviews). Products with high *vote* are products with high numbers of helpful votes per review (more trustworthy reviews).

For k-nearest neighbors (KNN), we trained our model in four different ways: using original data, using normalized data, using only top three variables, and using normalized top three variables. The corresponding MSE are $1.69 \times 10^{10}$, $1.69 \times 10^{10}$, $1.70 \times 10^{10}$ and $1.71 \times 10^{10}$. The corresponding correlations are 0.371, 0.381, 0.364 and 0.358. Normalized data with all variables produces the best KNN model. This makes sense because KNN involves calculating distances between data points, so data normalization is generally needed.

We also trained four other models using original data with all variables: gaussian naïve bayes, decision tree with max depth of 3, multilayer perceptron (MLP) and support vector regression. The corresponding MSE are $2.93 \times 10^{10}$, $1.73 \times 10^{10}$, $1.96 \times 10^{10}$ and $2.05 \times 10^{10}$. The corresponding correlations are 0.255, 0.351, 0.210 and 0.363. Interestingly, although a decision tree with max depth of 3 can only produce 8 distinct prediction values, it is arguably the best model among these four models. Unsurprisingly (or not), to navigate to the node with the best predicted rank, you need *also_buy_len* greater than 7.5, *also_view_len* greater than 17.5 and *price* greater than 2.34. The decision tree doesn't think highly of products that are too cheap.

We repeated the above analysis using only verified reviews to try to reduce the effect of fake reviews. The *verified* column now becomes a filter condition instead of a variable. After filtering, we reduced the number of data points from 896 to 669, which is still a good number of data points. Surprisingly, our models make worse predictions across the board when using only verified reviews, so it seems it is better to use *verified* as a variable. It is possible that the quality of verified reviews is not significantly higher than the quality of unverified reviews. Also, our models probably appreciate adding additional variables, because there are many data points but not that many variables.

We performed principal component analysis (PCA) on normalized data to create uncorrelated variables. We didn't reduce dimension because we only have a few variables. We retrained our models for random forest, KNN and OLS regression. Random forest becomes worse after PCA, KNN produces similar results after PCA, and OLS regression provides slightly better results after PCA. The improvement for OLS regression could be due to the fact that there is no correlation between principal components. Random forest doesn't seem to appreciate transformation of data.

When examining the residuals of our models, it is evident that the residuals are not normally distributed. Distribution of residuals tends to have fat right tail (positively skewed and high kurtosis), which is consistent with the fact that predicted rank rarely goes above 400000, but actual rank could reach 600000.

## Conclusion

We were able to build product rank prediction models with decent success, and all our variables have some predictive power. The best model is random forest trained on unnormalized data using all variables. The prediction accuracies are certainly not super impressive, and the distribution of residuals suggests that we might be missing other important variables in our models, or maybe some models are simply not appropriate for this dataset. For example, the independent variables might not have a linear relationship with the product rank, making linear regression inappropriate without data transformation. Other potentially important variables that are not covered by our analysis could include the brand of the product, the seller's rating, the sentiment of reviews, number of competing products, etc. Nevertheless, our model is able to make sensible predictions and is generally directionally correct. Based on our results, we would suggest people who want to sell their product on Amazon to design the webpage in a way that attracts attention, set the price lower than competitors, encourage buyers to leave 5-star reviews, and incentivize people to write thoughtful / helpful reviews.